\documentclass[11pt]{article}

\usepackage{amsmath}
\usepackage{graphicx}
\usepackage{amsfonts}
\usepackage{amssymb}
\usepackage{epsfig}
\usepackage{color}
\usepackage{psfrag}
\usepackage{epstopdf}
\usepackage{mathrsfs}

\newcommand{\EQ}{\begin{equation}}
\newcommand{\EN}{\end{equation}}
\newcommand{\be}{\begin{equation}}
\newcommand{\ee}{\end{equation}}
\newcommand{\bea}{\begin{eqnarray}}
\newcommand{\eea}{\end{eqnarray}}

\newcommand{\rd}{{\rm d}}

\usepackage{bm}
\usepackage{caption}
\definecolor{bluapple}{RGB}{0, 0, 255}
\definecolor{marroneapple}{RGB}{153, 102, 51}
\definecolor{cianoapple}{RGB}{0, 255, 255}
\definecolor{verdeapple}{RGB}{0, 255, 0}
\definecolor{magentaapple}{RGB}{255, 0, 255}
\definecolor{arancioapple}{RGB}{255, 127, 0}
\definecolor{violaapple}{RGB}{127, 0, 127}
\definecolor{rossoapple}{RGB}{255, 0, 0}
\definecolor{gialloapple}{RGB}{255, 255, 0}

\begin{document} %\setcounter{page}{0}
\topmargin 0pt
\oddsidemargin 5mm
\renewcommand{\thefootnote}{\arabic{footnote}}
\newpage
\topmargin 0pt
\oddsidemargin 5mm
\renewcommand{\thefootnote}{\arabic{footnote}}

\newpage

\begin{titlepage}
\begin{flushright}
%\color{red}{\boxed{\texttt{last update: July 26th, 2015}}}
\end{flushright}
%\vspace{0.5cm}
\begin{center}
{\large {\bf Multiple phases and vicious walkers in a wedge}}\\
%{\bf title}}\\
\vspace{1.8cm}
{\large Gesualdo Delfino$^\natural$ and Alessio Squarcini$^\flat$}\\ %\diamond
\vspace{0.5cm}
{\em SISSA -- Via Bonomea 265, 34136 Trieste, Italy}\\
{\em INFN -- sezione di Trieste, Italy}\\
	%{\em E-mail: delfino@sissa.it}\\
	%\vspace{0.5cm}
	%{\large and}\\
	%\vspace{0.5cm}
	%{\large P. Simonetti}\\
	%\vspace{0.5cm}
	%{\em Department of Physics, University of Wales Swansea,\\
	%Singleton Park, Swansea SA2 8PP, United Kingdom}\\
	%{\em email: p.simonetti@swansea.ac.uk}\\
\end{center}
\vspace{1.2cm}

\renewcommand{\thefootnote}{\arabic{footnote}}
\setcounter{footnote}{0}

\begin{abstract}
\noindent
We consider a statistical system in a planar wedge, for values of the bulk parameters corresponding to a first order phase transition and with boundary conditions inducing phase separation. Our previous exact field theoretical solution for the case of a single interface is extended to a class of systems, including the Blume-Capel model as the simplest representative, allowing for the appearance of an intermediate layer of a third phase. We show that the interfaces separating the different phases behave as trajectories of {\it vicious} walkers, and determine their passage probabilities. We also show how the theory leads to a remarkable form of wedge covariance, i.e. a relation between properties in the wedge and in the half plane, which involves the appearance of self-Fourier functions.
\end{abstract}

\vfill
$^\natural$delfino@sissa.it, $^\flat$alessio.squarcini@sissa.it
\end{titlepage}

%\tableofcontents
\newpage
%======================================================================================
%			SECTION1
%======================================================================================
\section{Introduction}
Fluid interfacial phenomena at boundaries form an important chapter of statistical physics and are studied experimentally, theoretically and numerically (see \cite{deGennes,Binder,Abraham_review,Diehl,Dietrich,Schick,FLN,BR,BLM,BEIMR} for reviews). On the theoretical side, the exact results obtained for the lattice Ising model in two dimensions \cite{Abraham_wetting,Abraham_strip,Abraham_review} provided an important benchmark for approximated or heuristic approaches, but proved too difficult to extend to other universality classes. Only recently it has been shown that phase separation and the interfacial region in planar systems can be described exactly for the different universality classes \cite{DV,DS2,DS1} relying on low energy properties of two-dimensional field theory \cite{fpu}. This new approach also allowed the exact solution \cite{DS3} of the longstanding problem of phase separation in a wedge, which received much attention \cite{Hauge,RDN,PRW,HKSD,APW,BCM,AM,RP,PR} as the basic example of the effect of the geometry of the substrate on the adsorption properties of a fluid. Particular interest was attracted by emergent relations between adsorption properties in the wedge and those on a flat substrate. Observed at the macroscopic level \cite{Hauge} and successively referred to as properties of ``wedge covariance'', this type of relations resisted a derivation within a statistical mechanical framework. It was one of the results of \cite{DS3} to show for the planar case how wedge covariance follows from the relativistic invariance of the quantum field theory associated to the universality class, which in turn reflects the homogeneous and isotropic nature of the fluid. 

The wedge problem has been considered so far for the case in which the universality class and the boundary conditions lead to the separation of two phases $a$ and $b$. Here we develop our exact field theoretical approach to study the case in which a macroscopic bubble of a third phase $c$ forms in between the two favored by the boundary conditions. Establishing whether a third phase will intrude between phases $a$ and $b$ forming a macroscopic intermediate layer or just microscopic droplets at the interface is a main question of wetting physics. It was shown in \cite{DV,DS2} that in two dimensions the answer is determined by the spectrum of elementary excitations of the underlying field theory, which is known for the different universality classes. A model which leads to the macroscopic (wetting) layer of the third phase and we will consider in this paper is the $q$-state Potts ferromagnet \cite{Wu} at its first order transition point, at which the $q$ ferromagnetic phases (two of which correspond to phases $a$ and $b$) coexist with the disordered phase (which will play the role of phase $c$). In two dimensions the first order transition corresponds to $T=T_c$ for $q>4$ \cite{Baxter}; strictly speaking, our field theoretical description is exact for the scaling limit $q\to 4^+$ \cite{DCq4}, but is expected to remain quantitatively meaningful up to $q\approx 10$, where the correlation length is still much larger than lattice spacing. Our derivation applies also to $q<4$ provided we allow for the possibility of vacant sites (dilution); then coexistence of the disordered with the ordered phases is recovered above a critical value of dilution for a value $T_c$ of the temperature. For $q=2$ one obtains a dilute Ising model, also known as Blume-Capel model, for which the wetting character of the disordered phase has been investigated numerically \cite{SY,SHK,AB}. 

We will work for values of the boundary parameters such that the inner phase is not adsorbed on the boundary. This means that the third phase is separated from the other two by two interfaces which fluctuate between the two boundary conditions changing points. We will see how this picture emerges within the field theoretical framework and will determine the passage probabilities for the interfaces, finding in particular that they randomly fluctuate with the constraint of avoiding each other and the boundary, i.e. that they correspond to trajectories of so-called ``vicious'' walkers \cite{Fisher}; in this way we determine the passage probabilities of vicious walkers in a wedge. Concerning the issue of wedge covariance, it turns out to acquire additional interest in the case of two interfaces, with a surprising interplay between physical considerations in momemtum space and mathematical realization of the condition of impenetrability of the wedge.

The derivations are exact and apply to the case of a shallow wedge, that for which the results are universal, in the sense that they do not depend on the specific values of the boundary parameters, as long these are in the range which does not bind the interfaces to the boundary.

The paper is organized as follows. In the next section we recall the setting and the results of \cite{DS3}. This will put us in the condition of developing the theory for the case of two interfaces in section~3. The final section is then devoted to summary and comments.

%======================================================================================
%			SECTION2
%======================================================================================
\section{Two phases in a wedge}
We start with the characterization of the statistical system in absence of boundaries, i.e. on the infinite plane. The system is considered at a first order phase transition point, where different phases, that we label by an index $a=1,2,\ldots,n$, have the same free energy and can coexist at equilibrium. At the same time the system is supposed to be close to a second order transition point\footnote{As an example, for the Ising ferromagnet these specifications amount to consider a temperature slightly below the critical value $T_c$, in absence of external field.}, in such a way that the correlation length is much larger than microscopic scales and a continuous description is allowed. For homogeneous and isotropic systems this continuous description is provided by a Euclidean field theory with coordinates $(x,y)$ identifying a point on the plane. This field theory in turn corresponds to the continuation to imaginary time $t=iy$ of a quantum field theory in one space dimension with coordinate $x$. The degenerate phases of the statistical system are in one-to-one correspondence with degenerate vacua $|0_a\rangle$ of the associated quantum theory. We denote by $\sigma(x,y)$ the order parameter field, and by $\langle\sigma\rangle_a=\langle 0_a|\sigma(x,y)|0_a\rangle$ the value of the order parameter in phase $a$. For a generic field $\Phi$ we have 
\EQ
\Phi(x,y) = \text{e}^{yH-ixP}\Phi(0,0)\text{e}^{-yH+ixP}\,,
\label{translations}
\EN
with the Hamiltonian $H$ and momentum operator $P$ of the quantum system acting as generators of time and space translations, respectively; the vacuum states carry zero energy and momentum.

As usual in presence of degenerate vacua in (1+1) dimensions (see \cite{fpu}), the elementary excitations correspond to kinks $\vert K_{ab}(\theta)\rangle$ which interpolate between two different vacua $\vert0_{a}\rangle$ and $\vert0_{b}\rangle$, and whose energy and momentum satisfy the relativistic dispersion relation 
\EQ
(e,p)=\left(m_{ab}\cosh\theta,m_{ab}\sinh\theta\right)\,,
\label{ep}
\EN 
where $m_{ab}$ is the kink mass (inversely proportional to the bulk correlation length) and $\theta$ is known as rapidity. Two vacua $\vert0_{a}\rangle$ and $\vert0_{b}\rangle$ (as well as the corresponding phases) are said to be adjacent if they can be connected by an elementary kink; when the connection requires a state $\vert K_{ac_{1}}(\theta_{1})K_{c_{1}c_{2}}(\theta_{2})\dots K_{c_{n-1}b}(\theta_{n})\rangle$, with $n$ necessarily larger than one, the two vacua are said to be non-adjacent.

As a further step towards the study of the wedge problem, we consider this statistical system on the half plane $x\geqslant0$. We call boundary condition of type $a$ a uniform (i.e. $y$-independent) boundary condition at $x=0$ favoring phase $a$ in the bulk\footnote{In a ferromagnet this is achieved applying a magnetic field on the boundary.}, in such a way that the order parameter approaches $\langle\sigma\rangle_a$ as $x\to +\infty$. We will use the notation $|0_a\rangle_0$ for the vacuum state of the quantum system on the half line with this boundary condition; more generally, the subscript $0$ will be used to indicate the presence of the vertical boundary. 

\begin{figure}[t]
\centering
\includegraphics[width=2.5cm]{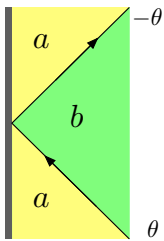}
\caption{A uniform boundary reflects a low energy kink.}
\label{rmatrix}
\end{figure}

Phase separation can be induced through a change of boundary conditions. Within the field theoretical description, the change of boundary conditions from type $a$ to type $b$ at a point $y$ on the boundary is realized by the insertion of a field $\mu_{ab}(0,y)$, with non-zero matrix elements on states interpolating between $|0_a\rangle_0$ and $|0_b\rangle_0$. When these two vacua are adjacent, which is the case we consider in this section, the simplest matrix element of $\mu_{ab}$ is\footnote{Here and below, in order to simplify the notation, we drop the indices on the kink mass.}
\EQ
{}_0\langle 0_a|\mu_{ab}(0,y)|K_{ba}(\theta)\rangle_0=e^{-my\cosh\theta}f_0(\theta)\,,
\label{bff}
\EN
where $f_0(\theta)$ gives the amplitude for the emission/absorption of a kink from the boundary condition changing point. The kink travels towards the boundary for $\theta<0$ (in-state), and away from it for $\theta>0$ (out-state). In- and out-states are related by the scattering operator \cite{ELOP}. As we are going to see, our computations involve low energy particles, whose scattering with the boundary is necessarily elastic, i.e. conserves the number of particles. Moreover, the field $\mu_{ab}$ acts on a uniform vertical boundary, which preserves the energy. For these reasons the low energy scattering of a particle on the boundary is a pure reflection (Fig.~\ref{rmatrix}), and the relation between the in- and out-state for (\ref{bff}) takes for small momenta the simple form $f_0(\theta)=\pm f_0(-\theta)$. On the other hand, only the choice
\EQ
f_0(\theta)=-f_0(-\theta)\,,\hspace{1cm}\theta\to 0\,
\label{unitarity1}
\EN
implies the property $f_0(0)=0$ which will eventually be responsible for the impenetrability of the wall. Generically, we will then have $f_0(\theta)=c_1\,\theta+O(\theta^2)$, with $c_1$ a constant.

Passing from a vertical boundary to one forming an angle $\psi$ with the vertical involves a rotation in Euclidean space, and then a relativistic transformation for the associated quantum field theory. Recalling (\ref{ep}), this transformation shifts rapidities by $i\psi$, so that the kink emission amplitude in the rotated frame, that we denote by $f_\psi$, is related to that in the original frame as
\EQ
f_\psi(\theta)=f_0(\theta+i\psi)\,;
\label{Fcovariance}
\EN
our considerations on $f_0$ then yield
\EQ
f_\psi(\theta)\simeq c_1(\theta+i\psi)\,,\hspace{1cm}|\theta|,\,|\psi|\ll 1\,.
\label{fmu}
\EN

\begin{figure}[t]
\centering
\includegraphics[width=3cm]{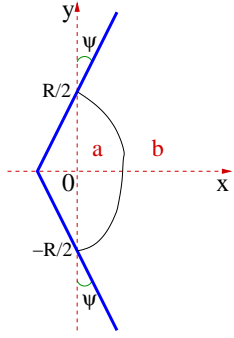}
\caption{Wedge geometry with boundary condition changing points at $(0,\pm R/2)$ and an interface running between them.}
\label{wedge_geometry2}
\end{figure}

At this point we are able to consider, instead of the half plane, the more general wedge geometry of Fig.~\ref{wedge_geometry2}. The points $(0,\pm R/2)$ are boundary condition changing points, such that phase $b$ (resp. $a$) is favored for $|y|>R/2$ (resp. $|y|<R/2$) on the wedge. For $mR$ large, i.e. when the system is observed on a scale much larger than the bulk fluctuations, one then expects an interface running between the points $(0,\pm R/2)$, separating an inner phase $a$ from an outer phase $b$. These expectations emerge from the theory in the following way. For $|y|<R/2$ the order parameter in the wedge, that we denote by $\langle\sigma(x,y)\rangle_{W_{bab}}$, reads
\EQ
\langle \sigma(x,y) \rangle_{W_{bab}} = \frac{{}_\psi\langle 0_{b} \vert \mu_{ba}(0,\frac{R}{2})\sigma(x,y)\mu_{ab}(0,-\frac{R}{2}) \vert 0_{b} \rangle_{-\psi}}{{Z}_{W_{bab}}}\,, 
\label{op}
\EN
where the subscripts $\pm\psi$ indicate the different rotations performed for positive and negative $y$, and 
\bea
{Z}_{W_{bab}} &=& {}_\psi\langle 0_{b} \vert \mu_{ba}(0,R/2)\mu_{ab}(0,-R/2) \vert 0_{b} \rangle_{-\psi}\,\nonumber\\
&\sim & \int_0^\infty\frac{\rd \theta}{2\pi}\,f_\psi(\theta)f_{-\psi}(\theta)\,\text{e}^{-mR(1+\frac{\theta^2}{2})}
\sim \frac{c_1^{2}\,\text{e}^{-mR}}{2\sqrt{2\pi}(mR)^{3/2}} \, (1 + mR \,\psi^2)\,.
\label{partition02}
\eea
This result is obtained inserting a complete set of particle states in between the two fields, taking the limit $mR$ large which projects on the lightest (single-kink) intermediate state and to small rapidities, and considering $\psi$ small in order to use (\ref{fmu}); here and in the following the symbol $\sim$ indicates omission of terms subleading for $mR$ large. In a similar way we obtain
\bea \nonumber
\langle \sigma(x,y) \rangle_{W_{bab}} &\sim &\frac{e^{-mR}}{{Z}_{W_{bab}}} \int_{-\infty}^{+\infty}\frac{\rd \theta_{1} \rd \theta_{2}}{(2\pi)^{2}}
\text{e}^{-\frac{m}{2}[(\frac{R}{2}-y)\theta_{1}^2+(\frac{R}{2}+y)\theta_{2}^2]-imx(\theta_{1}-\theta_{2})}\\
& \times & f_{\psi}(\theta_{1})\, \langle K_{ba}(\theta_{1})\vert \sigma(0,0) \vert K_{ab}(\theta_{2}) \rangle\,f_{-\psi}(\theta_{2})\,,
\label{op1}
\eea
where we evaluate the order parameter field on bulk states, implying that the boundary condition changing fields account for the leading boundary effects at large $R$. The matrix element in (\ref{op1}) contains a disconnected part proportional to $\delta(\theta_1-\theta_2)$ which yields a constant after integration, and then does not contribute to the derivative with respect to $x$ we are going to take in a moment; the behavior of the connected part in the relevant region $\theta_1,\theta_2\to 0$, is instead determined by the `kinematical' pole (see \cite{fpu} and references therein)
\EQ
\langle K_{ba}(\theta_{1})\vert \sigma(0,0) \vert K_{ab}(\theta_{2}) \rangle_{\textrm{connected}}\simeq i\frac{\langle\sigma\rangle_b-\langle\sigma\rangle_a}{\theta_1-\theta_2}\,,\hspace{1cm}\theta_1\simeq\theta_2\,.
\label{kinematical}
\EN
With this information we obtain \cite{DS3}
\EQ
\frac{\partial_{x}\langle \sigma(x,y)\rangle_{W_{bab}}}{\langle\sigma\rangle_b-\langle\sigma\rangle_a}\sim 8\sqrt{2}\left( \frac{m}{R}\right)^{\frac{3}{2}} \frac{\left(x+\frac{R\psi}{2}\right)^{2}-(\psi y)^{2}}{\sqrt{\pi}\,\kappa^{3}(1+mR\psi^{2})}\,\text{e}^{-\chi^{2}},
\label{dop}
\EN
where
\EQ
\kappa=\sqrt{1-\epsilon^2}\,,\hspace{1cm}\epsilon=\frac{2y}{R}\,,\hspace{1cm}\chi=\sqrt{\frac{2m}{R}}\,\frac{x}{\kappa}\,,
\label{defs1}
\EN
and, integrating back over $x$ with the condition $\langle\sigma(+\infty,y)\rangle_{W_{bab}}=\langle\sigma\rangle_{b}$,
\EQ
\langle \sigma(x,y) \rangle_{W_{bab}}\sim\langle\sigma\rangle_{a}+[\langle\sigma\rangle_b-\langle\sigma\rangle_a]\Biggl[\text{erf}(\chi)
-\frac{2}{\sqrt{\pi}}\,\frac{\chi + \sqrt{2mR}\,\frac{\psi}{\kappa}}{1+mR\psi^{2}}\,\text{e}^{-\chi^{2}} \Biggr];
\label{op2}
\EN
for $\psi=y=0$ and $\langle\sigma\rangle_{a}=-\langle\sigma\rangle_{b}$ this result coincides with that obtained in \cite{Abraham_wetting} from the lattice solution of the Ising model on the half plane. 

\begin{figure}[t]
\centering
\includegraphics[width=7cm]{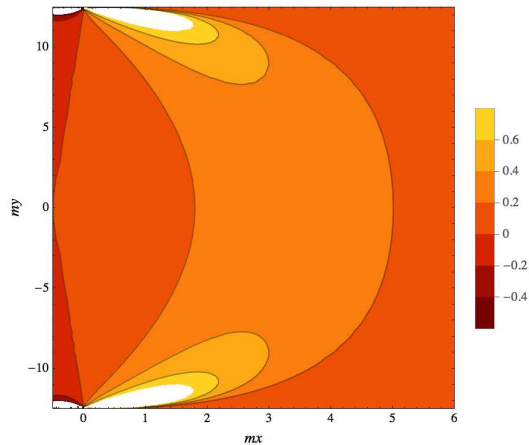}
\caption{Contour plot of the passage probability density $P_1^{(\psi)}(x;y)/m$ for $mR=25$, $\psi=0.04$. The leftmost contour line corresponds to $P_1^{(\psi)}(x;y)=0$, and then to the wedge.}
\label{wedge_pp}
\end{figure}

It was shown in \cite{DV,DS2} that the leading large $R$ contribution to the order parameter profile, i.e. the one associated to the pole in (\ref{kinematical}), corresponds to a sharp phase separation between pure phases. In the present case of adjacent phases, there will be a single interface, with a probability $P^{(\psi)}_1(x;y)$ to intersect the line of constant ordinate $y$ in the interval $(x,x+dx)$. It follows that the leading large $R$ expression of the order parameter can be written as 
\EQ
\langle \sigma(x,y) \rangle_{W_{bab}}\sim\langle \sigma \rangle_{b} \int_{\tilde{x}}^{x} \rd u \, P_1^{(\psi)}(u;y) + \langle \sigma \rangle_{a} \int_{x}^{\infty} \rd u \, P_1^{(\psi)}(u;y)\,,
\label{prob1}
\EN
where $\tilde{x}(y)$ is the abscissa of the point on the wedge with ordinate $y$; this expression shows that $P_1^{(\psi)}(x;y)$ actually coincides with (\ref{dop}). Also for later use we introduce the additional notations
\EQ
\lambda=\sqrt{\frac{R}{2m}}\,,\hspace{1cm}\eta=\frac{x}{\lambda}\,,\hspace{1cm}
\hat{\psi}=\sqrt{\frac{mR}{2}}\,\psi\,,
\label{defs2}
\EN
and rewrite this result as
\EQ
P_1^{(\psi)}(x;y)\sim\frac{4}{\sqrt{\pi}\kappa^3\lambda(1+2\hat{\psi}^2)}[(\eta+\hat{\psi})^2-(\epsilon\hat{\psi})^2]e^{-\chi^2}\,.
\label{P1}
\EN
The requirement $\int_{\tilde{x}}^{\infty}\rd x\,P_1^{(\psi)}(x;y)\approx 1$ is satisfied as long as $\sqrt{mR}\,\psi\ll 1$. Notice that (\ref{dop}) or (\ref{P1}) show that $P_1^{(\psi)}(x;y)$ vanishes for $|y|=\frac{x}{\psi}+\frac{R}{2}$, which for the present case of small $\psi$ are the coordinates of the wedge ($x\geq-R\psi/2$); hence, the properties (\ref{Fcovariance}), (\ref{fmu}) that we identified in momentum space indeed lead to an impenetrable wedge in coordinate space. A plot of $P_1^{(\psi)}(x;y)$ is shown in Fig.~\ref{wedge_pp}. 

%======================================================================================
%			SECTION3
%======================================================================================
\section{Third phase and double interface}
In this section we still consider the wedge geometry of Fig.~\ref{wedge_geometry2} with the same boundary conditions $bab$ of the previous section, but now we study the case in which the phases $a$ and $b$ are not adjacent. More precisely, we consider the case in which the lightest state connecting $\vert 0_{a} \rangle$ and $\vert 0_{b} \rangle$ is the two-kink state $|K_{ac}(\theta_1)K_{cb}(\theta_2)\rangle$, with a unique choice of the intermediate vacuum $|0_c\rangle$. This situation arises, in particular, in the (dilute) $q$-state Potts model at first order transition that we discussed in the introduction. Indeed, the model is exactly solvable (integrable) in the scaling limit, and it is known  that there are no kinks directly connecting two ferromagnetic vacua at the first order transition \cite{dilute,DCq4}. The lightest state connecting two such vacua $|0_a\rangle$ and $|0_b\rangle$ is the two-kink state $|K_{a0}K_{0b}\rangle$ passing by the disordered vacuum $|0_0\rangle$; the symmetry under permutations of the $q$ ferromagnetic phases which characterizes the Potts model \cite{Wu} ensures that the elementary kinks $K_{a0}$, $K_{0a}$ ($a=1,\ldots,q$) all have the same mass $m$. 

Technically, the difference with respect to the previous section is that now the large $R$ expansion of (\ref{op}) is dominated by the contribution of the two-kink state. In particular, the relevant matrix element for the boundary condition changing fields is no longer (\ref{bff}) but
\EQ
{}_0\langle 0_a|\mu_{ab}(0,y)|K_{bc}(\theta_1)K_{ca}(\theta_2)\rangle_0=e^{-my(\cosh\theta_1+\cosh\theta_2)}f_0(\theta_1,\theta_2)\,.
\label{bff2}
\EN
As before, the low energy scattering properties of the kinks on a vertical wall can be used to infer properties of the amplitude $f_0(\theta_1,\theta_2)$, but now we will also exploit the integrability of the scaling (dilute) Potts model at the first order transition. Integrability ensures that the interaction of the two kinks on the wall can be regarded as consisting of two independent processes (factorization of the scattering \cite{GZ}), and this in turn allows us to write a relation like (\ref{unitarity1}) for each particle, i.e.
\EQ
f_0(\theta_1,\theta_2)=-f_0(-\theta_1,\theta_2)=-f_0(\theta_1,-\theta_2)\,,
\hspace{1cm}\theta_1,\theta_2\to 0\,.
\label{unitarity2}
\EN
Integrability also yields the exact bulk scattering matrix of the scaling (dilute) Potts model at the first order transition \cite{dilute,DCq4}. From this one reads, in particular, that at low energy the state $|K_{b0}(\theta_1)K_{0a}(\theta_2)\rangle$ scatters in the bulk into the state $-|K_{b0}(\theta_2)K_{0a}(\theta_1)\rangle$, so that we have the additional relation
\EQ
f_0(\theta_1,\theta_2)=-f_0(\theta_2,\theta_1)\,,\hspace{1cm}
\theta_1,\theta_2\to 0\,.
\label{unitarity3}
\EN
Equations (\ref{unitarity2}) and (\ref{unitarity3}) lead to
\EQ
f_0(\theta_1,\theta_2)\simeq c_2\,\theta_1\theta_2(\theta_1^2-\theta_2^2)
\,,\hspace{1cm}\theta_1,\theta_2\ll 1\,.
\label{f0}
\EN
As before, the passage from the vertical boundary to that rotated by an angle $\psi$ involves a rapidity shift,
\EQ
f_\psi(\theta_1,\theta_2)=f_0(\theta_1+i\psi,\theta_2+i\psi)\,,
\label{f2psi}
\EN
and the leading large $mR$ expression for the order parameter in the wedge can be written as\footnote{Generically, we keep the notation $c$ for the third phase; $c=0$ for the Potts case.}
\bea\nonumber
\langle\sigma(x,y)\rangle_{W_{bab}} & \sim & \frac{1}{{Z}_{W_{bab}}}\int_{\mathbb{R}^{4}} \frac{\rd\theta_{1}\rd\theta_{2}\rd\theta_{3}\rd\theta_{4}}{(2\pi)^{4}} \, f_\psi(\theta_{4},\theta_{3})\langle K_{bc}(\theta_{3})K_{ca}(\theta_{4}) \vert \sigma(0,0) \vert K_{ac}(\theta_{1})K_{cb}(\theta_{2}) \rangle\\
& \times & f_{-\psi} (\theta_{1},\theta_{2})\mathcal{Y}(\theta_{1},\theta_{2},\theta_{3},\theta_{4};x,y)\,,
\label{op3}
\eea
where
\EQ
\mathcal{Y}(\theta_{1},\theta_{2},\theta_{3},\theta_{4};x,y) = U^{-}(\theta_{1};x,y)U^{-}(\theta_{2};x,y)U^{+}(\theta_{3};x,y)U^{+}(\theta_{4};x,y)\,,
\EN
\EQ
U^{\pm}(\theta;x,y) = \text{e}^{m\left(-\frac{R}{2}\pm y\right)\cosh\theta\mp imx\sinh\theta}\,,
\EN
\bea
{Z}_{W_{bab}} &\sim &\int_{\mathbb{R}_{+}^{2}} \frac{\rd\theta_{1}\rd\theta_{2}}{(2\pi)^{2}} \, f_\psi(\theta_{2},\theta_{1})f_{-\psi} (\theta_{1},\theta_{2})\,\text{e}^{-\frac{mR}{2}\left(\cosh\theta_{1}+\cosh\theta_{2}\right)}\nonumber\\
&\sim & \frac{3c_2^2}{\pi} \frac{\text{e}^{-2mR}}{(mR)^{5}}\left[
1 + \left(8-\frac{32}{3\pi}\right)\hat{\psi}^{2}+O(\hat{\psi}^4)\right]\,.
\label{zeta2}
\eea

\begin{figure}[t]
\centering
\includegraphics[width=9cm]{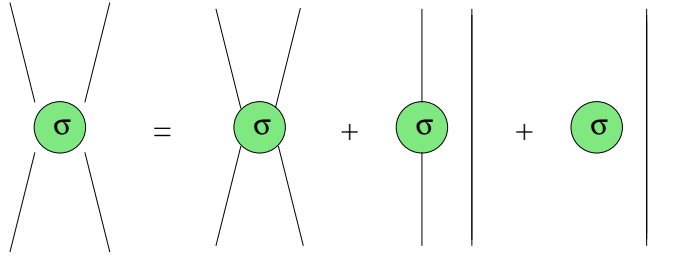}
\caption{The four-leg matrix element of the order parameter field $\sigma$ decomposes into the sum of the connected and disconnected parts in the r.h.s.}
\label{four_leg}
\end{figure}

The matrix element of the field $\sigma$ on two-kink states entering (\ref{op3}) contains three types of contributions, depending on the number of annihilations that arise when particles on the left and on the right have the same rapidity. The three contributions are schematically depicted in Fig.~\ref{four_leg} and correspond to a connected part (no annihilations), a partially disconnected part (one annihilation) and a totally disconnected part (two annihilations). The complete computation of the order parameter taking into account all these contributions has been performed in \cite{DS2} for the case of the strip geometry and in \cite{thesis} for the case of the half plane ($\psi=0$). As expected from the fact that the two-kink state yields the leading contribution, the results correspond to the presence of two interfaces separating the intermediate phase $c$ from the phases $a$ and $b$. For the case of the wedge the complete calculation becomes cumbersome, and it is particularly interesting that we can still obtain the complete results in a relatively simple way through the following procedure, whose exactness we explicitly checked for the strip and the half plane. 

Generalizing what seen in the previous section, the probability $P_2^{(\psi)}(x_1,x_2;y)$ that one interface intersects the line of constant ordinate $y$ in the interval $(x_1,x_1+dx)$, and that the other interface intersects the same line in the interval $(x_2,x_2+dx)$ is related to the order parameter as
\begin{equation}
\label{probabilisticprofile}
\langle\sigma(x,y)\rangle_{W_{bab}} = \int_{\tilde{x}}^{+\infty}\rd x_{1} \, \int_{\tilde{x}}^{+\infty}\rd x_{2} \, P_{2}^{(\psi)}(x_{1},x_{2};y) \sigma(x\vert x_{1},x_{2})\,,
\end{equation}
where
$$
\sigma(x\vert x_{1},x_{2}) = \left\{
\begin{array}{ll}
\langle\sigma\rangle_{a}\,,	& \tilde{x}<x < \text{min}(x_{1},x_{2})\,, \\
\langle\sigma\rangle_{c}\,,	& \text{min}(x_{1},x_{2}) < x < \text{max}(x_{1},x_{2})\,,\\
\langle\sigma\rangle_{b}\,,	& x > \text{max}(x_{1},x_{2})\,, \\
\end{array}
\right.
$$
and $\tilde{x}(y)$ is the abscissa of the wedge. On the other hand 
\EQ
{P}_{1,2}^{(\psi)}(x_1;y)=\int_{\tilde{x}}^{\infty}\rd x_{2}\, P_{2}^{(\psi)}(x_{1},x_{2};y)
\label{P12}
\EN
is the probability that one of the two interfaces passes in the interval $(x_1,x_1+dx)$ at ordinate $y$, irrespectively of the other. Since it is the field $\sigma$ which `detects' the interfaces, it is natural to expect, and we checked explicitly that this is the case for the strip and the half plane, that ${P}_{1,2}^{(\psi)}(x;y)$ is determined by the second term in the r.h.s. of Fig.~\ref{four_leg}, proportional to\footnote{Of course there are analogous terms with different pairings of rapidities, all giving the same contribution to (\ref{op3}).} 
\EQ
\langle K_{ba}(\theta_{3})\vert \sigma(0,0) \vert K_{ab}(\theta_{1}) \rangle_{\textrm{connected}}\delta(\theta_2-\theta_4)\,,
\label{2leg}
\EN
and to which we refer as the two-leg term, from the number of particles connected to the field $\sigma$. Up to the factor $\delta(\theta_2-\theta_4)$, corresponding to the undetected interface, this two-leg term is the same we studied in the previous section for the single interface. Plugging (\ref{2leg}) into (\ref{op3}) we obtain
\bea\nonumber
\langle\sigma(x,y)\rangle_{W_{bab}}^{\textrm{two-leg}} & \propto & \int_{\mathbb{R}^{4}}\rd\theta_{1}\rd\theta_{2}\rd\theta_{3}\rd\theta_{4}\, f_\psi(\theta_{4},\theta_{3})\frac{\delta(\theta_2-\theta_4)}{\theta_1-\theta_3}f_{-\psi} (\theta_{1},\theta_{2})\\
& \times & e^{-m\left[\frac{R}{4}\sum_{k=1}^4\theta_k^2+\frac{y}{2}(\theta_1^2+\theta_2^2-\theta_3^2-\theta_4^2)-ix(\theta_1+\theta_2-
\theta_3-\theta_4)\right]}\,,
\label{op4}
\eea
where, as usual, we took into account that small rapidities dominate at large $R$ and we used (\ref{kinematical}); as in the previous section, a derivative with respect to $x$ cancels the pole. On the other hand, we can also write $\langle\sigma(x,y)\rangle_{W_{bab}}^{\textrm{two-leg}}$ in a way analogous to (\ref{prob1}), with ${P}_{1,2}^{(\psi)}$ replacing ${P}_{1}^{(\psi)}$, and $\langle\sigma\rangle_c$ replacing $\langle\sigma\rangle_a$ or $\langle\sigma\rangle_b$. It follows that ${P}_{1,2}^{(\psi)}(x;y)\propto\partial_x\langle\sigma(x,y)\rangle_{W_{bab}}^{\textrm{two-leg}}$, i.e.
\bea\nonumber
{P}_{1,2}^{(\psi)}(x_1;y) & \propto & \int\rd x_2\int_{\mathbb{R}^{4}}\rd\theta_{1}\rd\theta_{2}\rd\theta_{3}\rd\theta_{4}\, f_\psi(\theta_{4},\theta_{3})f_{-\psi} (\theta_{1},\theta_{2})\\
& \times & e^{-m\left[\frac{R}{4}\sum_{k=1}^4\theta_k^2+\frac{y}{2}(\theta_1^2+\theta_2^2-\theta_3^2-\theta_4^2)+ix_1(\theta_1-\theta_3)-ix_2(\theta_2-\theta_4)\right]}\,,
\label{P122}
\eea
where we used $\delta(z)\propto\int\rd s\,e^{isz}$. Comparison with (\ref{P12}) shows that the integrand of the integral in $x_2$ in (\ref{P122}) is proportional to ${P}_{2}^{(\psi)}$. Since the identity
\begin{equation}
\int_{\mathbb{R}^{2}}\rd\beta_{1}\rd\beta_{2} \,f_{\psi}(\beta_{1},\beta_{2}) \text{e}^{-\frac{\beta_{1}^{2}+\beta_{2}^{2}}{2}+iq_{1}\beta_{1}+iq_{2}\beta_{2}} = 2\pi f_{-i\psi}(q_{1},q_{2}) \text{e}^{-\frac{q_{1}^{2}+q_{2}^{2}}{2}} ,
\label{fourier}
\end{equation}
holds for the function defined by (\ref{f0}) and (\ref{f2psi}), we finally obtain
\bea
{P}_{2}^{(\psi)}(x_1,x_2;y) &=& \frac{N_{\hat{\psi}}}{\lambda^2\kappa^{10}}\,f_{-i(1+\epsilon)\hat{\psi}}(\eta_{1},\eta_{2})f_{-i(1-\epsilon)\hat{\psi}}(\eta_{1},\eta_{2}) \text{e}^{-\chi_{1}^{2}-\chi_{2}^{2}}
\label{P2}\\
& = & \frac{N_{\hat{\psi}}}{\lambda^2\kappa^{10}}\,f_{0}(\eta_{1}+(1+\epsilon)\hat{\psi},\eta_{2}+(1+\epsilon)\hat{\psi})\,f_{0}(\eta_{1}+(1-\epsilon)\hat{\psi},\eta_{2}+(1-\epsilon)\hat{\psi}) \text{e}^{-\chi_{1}^{2}-\chi_{2}^{2}}\,,\nonumber
\eea
where we are using the notations (\ref{defs1}) and (\ref{defs2}) with $\eta_i$ and $\chi_i$ corresponding to $x_i$, and $N_{\hat{\psi}}$ is dimensionless and determined by the condition $\int_{\tilde{x}}^\infty\rd x P_{1,2}(x;y)=1$. 

\begin{figure}
  \centering
  \includegraphics[width=9cm]{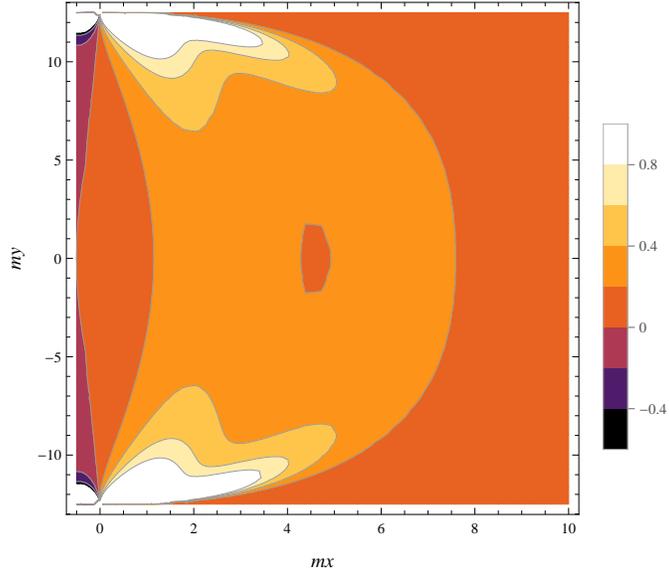}
\caption{Contour plot of the passage probability density $m^{-1}{P}_{1,2}^{(\psi)}(x;y)$ for $mR=25$ and $\psi=0.04$. The leftmost contour corresponds to ${P}_{1,2}^{(\psi)}=0$, and then to the wedge.}
\label{densitywedge}
\end{figure}

Recalling the form (\ref{f0}) of the function $f_0$, we see that the joint passage probability density (\ref{P2}) factors the terms $(\eta_{1}-\eta_{2})^{2}$ and $\pm\epsilon=1+\eta_i/\hat{\psi}$ (i.e. $\pm y = \frac{R}{2} + \frac{x_i}{\psi}$). It follows that the considerations in momentum space that led us to the result (\ref{f2psi}) for the function $f_\psi(\theta_1,\theta_2)$ produce in coordinate space a mutual repulsion among the interfaces (${P}_{2}^{(\psi)}(x,x;y)=0$), as well as the presence of an impenetrable wedge along which the passage probability density vanishes. A plot of ${P}_{1,2}^{(\psi)}(x;y)$ is shown in Fig.~\ref{densitywedge}.

For $\psi=0$ (\ref{P2}) reduces to
\begin{equation}
\label{passagedouble}
P_{2}^{(0)}(x_{1},x_{2};y) = \frac{16}{3\pi}\frac{\chi_{1}^{2}\chi_{2}^{2}\left(\chi_{1}^{2}-\chi_{2}^{2}\right)^{2}}{\kappa^{2}\lambda^{2}} \text{e}^{-\chi_{1}^{2}-\chi_{2}^{2}}\,,
\end{equation}
a result which is known \cite{Schehr_et_al} to correspond to the so-called ``vicious'' walkers \cite{Fisher} on the half line $x\geq 0$: the walkers start at $x=0$, move randomly with the constraint of avoiding each other and the boundary, and return to $x=0$ after a time $R$. Hence, our result (\ref{P2}) yields the exact joint passage probability density of two vicious walkers in the wedge.

\begin{figure}
\centering
\includegraphics[width=9cm]{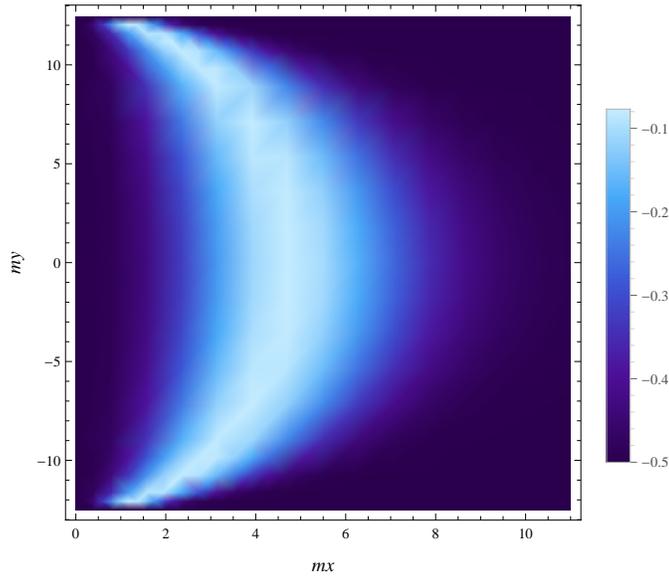}
\caption{Order parameter $M^{-1}\langle\sigma_{3}(x,y)\rangle_{W_{121}}^{\psi=0}$ and contour lines for the dilute three-state Potts model at first order transition. Due to permutational symmetry, $\sigma_3$ does not distinguish between phases 1 and 2, and the intermediate bubble of the disorderd phase is clearly visible.}
\label{tricritical}
\end{figure}

\begin{figure}[t]
\centering
\includegraphics[width=10cm]{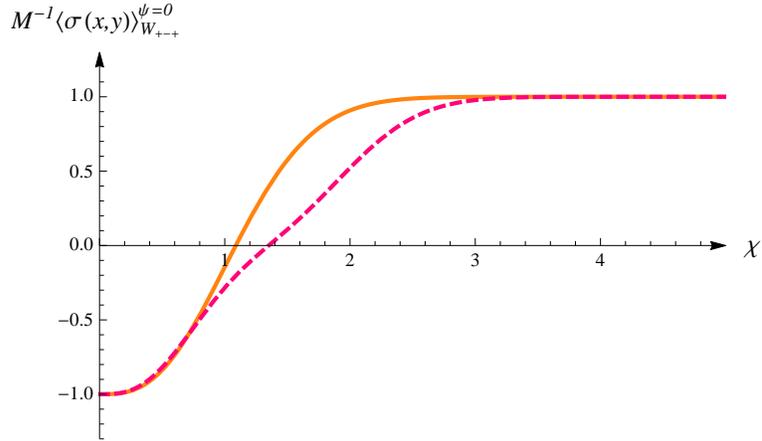}
\caption{Order parameter $M^{-1}\langle\sigma(x,y)\rangle_{W_{+-+}}^{\psi=0}$ for the Ising model in the undilute (continuous) and dilute (dashed) cases; in the dilute case the intermediate disordered bubble smoothens the profile.}
\label{dilute_undilute}
\end{figure}

The order parameter can be determined from $P_2^{(\psi)}$ through (\ref{probabilisticprofile}). We quote here the explicit result in the case of the half plane, for which the expressions simplify. For the Potts model the order parameter field has components $\sigma_k$ ($\sum_{k=1}^q\sigma_k=0$), with $\langle\sigma_k\rangle_0=0$ in the disordered phase, and we obtain\footnote{These results coincide with those derived by direct summation of all terms in Fig.~\ref{four_leg} \cite{thesis}.}
\begin{equation}
\label{dilutemag}
\langle\sigma_{k}(x,y)\rangle_{W_{bab}}^{\psi=0} = \left[\langle\sigma_{k}\rangle_{b}+\langle\sigma_{k}\rangle_{a}\right]{\cal A}(\chi) -2\langle\sigma_{k}\rangle_{a} {\cal B}(\chi) + \langle\sigma_{k}\rangle_{a}\,,
\end{equation}
with 
\EQ
\label{vev1}
\langle \sigma_{k} \rangle_{a} = \frac{q\delta_{ka}-1}{q-1}M\,,\hspace{.8cm}a,k=1,\ldots q\,,
\EN
\begin{equation}
{\cal A}(\chi) = -\frac{4}{3\pi}\chi^{2}\left(\chi^2-3\right)\text{e}^{-2\chi^2} + \frac{2}{3\sqrt{\pi}}\chi\left(-2 \chi ^4+\chi ^2-6\right)\text{e}^{-\chi^2} \text{erf}(\chi) + \text{erf}(\chi )^2\,,
\end{equation}
\begin{equation}
{\cal B}(\chi) = \frac{\chi}{3\sqrt{\pi}}\left(-6+\chi^{2}-2\chi^{4}\right) \text{e}^{-\chi^{2}} + \text{erf}(\chi)\,.
\end{equation}
A plot is shown in Fig.~\ref{tricritical} for $q=3$; for $q=2$ (Ising) Fig.~\ref{dilute_undilute} compares the order parameter profile in the dilute case with the undilute result of Ref.~\cite{DS2}.

\section{Conclusion}
In this paper we developed the theory of phase separation in a planar wedge for the case in which a macroscopic bubble of a third phase forms in between the two phases favored by the boundary conditions. We discussed the $q$-state Potts model (dilute for $q<4$) at its first order transition as an example to which the theory applies. In principle the full field theoretical calculation is much more complicated than that performed in \cite{DS3} for the case of a single interface. However, we found that, isolating a specific contribution to the order parameter which corresponds to the detection of a single interface, the formalism allows to reconstruct the complete result. This finding, that we checked explicitly against the full calculations for the cases of the strip and of the half plane, appears promising for further developments.

For the case of the wedge, it is worth stressing that the very fact that the final result exhibits a wedge-shaped path along which the passage probability for the interfaces vanishes provides a non-trivial consistency check for the theory. Indeed, the calculation starts from considerations in momentum space, in which the presence of a boundary is codified in properties of matrix elements of boundary condition changing fields. Moreover, the boundary is initially flat, and the 
information about the wedge is introduced through relativistic transformations performed on the matrix elements, always in momentum space. While in principle these are the same logic steps performed in \cite{DS3} for the separation of two phases, in practice the present case with a third phase is much more structured and leads to the specific form (\ref{f0}), (\ref{f2psi}) for the matrix element $f_\psi(\theta_1,\theta_2)$ of the boundary condition changing field. It is then remarkable to realize going on with the computation that the wedge in real space emerges because $f_\psi(\theta_1,\theta_2)$ turns out to fulfill the self-Fourier transform property (\ref{fourier}). In this way, the mechanism which eventually accounts for wedge covariance acquires surprising mathematical implications in presence of a third phase.

This appears to have also additional implications. Indeed, we arrived at (\ref{f0}) exploiting also the integrability of the scaling Potts model, which in turn ensures factorization of the scattering and equation (\ref{unitarity2}). On the other hand, since (\ref{unitarity2}) is necessary to arrive at (\ref{f0}), and the latter is necessary for the appearance of the wedge in real space through (\ref{fourier}), we are led to conclude that factorization of the scattering {\it at low energies} is required in systems allowing for the appearance of a bubble of a third phase. Notice that this is a weaker property than integrability, which implies factorization of the scattering at all energies.

We determined the joint passage probability for the interfaces separating the three phases and found that, in the case of a flat boundary (tilt angle $\psi=0$ for the wedge), it coincides with the known probability for vicious walkers in the half plane. Hence for $\psi\neq 0$ our result provides the passage probability for vicious walkers in a wedge. The name vicious walkers is used in the literature for random walkers subject to the constraint of avoiding each other and the boundary. In our framework the properties of random propagation and avoidance for the interfaces emerge from the limit of large separation $R$ between the boundary condition changing points on the boundary (pinning points for the interfaces), which is needed to observe phase separation. This limit projects the dynamics of the particles to low energies, where it turns out to reduce to fermionic statistics and becomes universal. Consistently, the average distances between the interfaces and between the interfaces and the boundary grow as $\sqrt{R/m}$, and are much larger than the range $1/m$ of the particle-particle and particle-boundary interactions, whose details then affect only subleading orders in the large $mR$ expansion.

\vspace{1cm}
\section*{Acknowledgments}
\noindent We thank the Galileo Galilei Institute for Theoretical Physics in Arcetri for hospitality during the program Statistical Mechanics, Integrability and Combinatorics, where part of this work was carried out.

%\newpage

\end{document}